\documentclass[aps,prl,preprint,groupedaddress]{revtex4}
\usepackage{graphicx}
\usepackage[english]{babel}

\def\mycmd{1}
\newcommand{\Hide}[1]{
\ifx\mycmd\undefined
Figure command is not define!
\else
  \if\mycmd1
  #1
  \else
 Hidden Figure!
  \fi
\fi}

\begin{document}

\title{
Time-Resonant Tokamak  Plasma Edge Instabilities?}

\author{A. J. W\fontshape{sc}\selectfont{ebster}$^{1,2}$}
\author{R. O. D\fontshape{sc}\selectfont{endy}$^{1,2,3}$}
\author{F. A. C\fontshape{sc}\selectfont{alderon}$^{1,3}$}
\author{S. C. C\fontshape{sc}\selectfont{hapman}$^{1,3}$}
\author{E. D\fontshape{sc}\selectfont{elabie}$^{1,4}$}
\author{D. D\fontshape{sc}\selectfont{odt}$^{1,2,5}$}
\author{R. F\fontshape{sc}\selectfont{elton}$^{1,2}$}
\author{T. N. T\fontshape{sc}\selectfont{odd}$^{1,2}$}
\author{F. M\fontshape{sc}\selectfont{aviglia}$^{1,2}$}
\author{J. M\fontshape{sc}\selectfont{orris}$^{1,2}$}
\author{V. R\fontshape{sc}\selectfont{iccardo}$^{1,2}$}
\author{B. A\fontshape{sc}\selectfont{lper}$^{1,2}$}
\author{S. B\fontshape{sc}\selectfont{rezinsek}$^{1,6}$}
\author{P. C\fontshape{sc}\selectfont{oad}$^{1,2}$}
\author{J. L\fontshape{sc}\selectfont{ikonen}$^{1,7}$}
\author{M. R\fontshape{sc}\selectfont{ubel}$^{1,8}$}
\author{JET EFDA Contributors\footnote{See the Appendix of
    F. Romanelli et al., Proceedings of the 24th IAEA Fusion Energy
    Conference 2012, San Diego, US.}}
\affiliation{$^1$JET-EFDA, Culham Science Centre, Abingdon, OX14 3DB, UK}
\affiliation{$^2$ EURATOM/CCFE Fusion Association,
  Culham Science 
  Centre, Abingdon, OX14 3DB, UK}
\affiliation{$^3$ Centre for Fusion, Space and Astrophysics, Department of
  Physics, University of Warwick, Coventry, CV4 7AL, UK}
\affiliation{$^4$ FOM Institute DIFFER, Association EURATOM-FOM,
  Nieuwegein, The Netherlands.}
\affiliation{$^5$ Max Planck Institut f\"ur Plasmaphysik, EURATOM
  ASSOCIATION, D-85748 Garching, Germany.}
\affiliation{$^6$ IEK-Plasmaphysik, Forschungszentrum Jülich,
  Association EURATOM-FZJ, Jülich, Germany.}  
\affiliation{$^7$ VTT, Association Euratom-Tekes, PO Box 1000,
  FI-02044 VTT, Finland.}  
\affiliation{$^8$ Alfv\'en Laboratory, School of Electrical Engineering,
  Royal Institute of Technology (KTH), Association EURATOM-VR,
  Stockholm, Sweden.}


\date{\today}

\begin{abstract}

For a two week period during the Joint European Torus
(JET)  2012 experimental campaign, the same high confinement plasma
was repeated 151 times. 
The dataset was analysed to produce a probability density function
(pdf) for the waiting times between edge-localised plasma
instabilities (``ELMs'').  
The result was entirely unexpected. 
Instead of a smooth single peaked
pdf, a succession of 4-5 sharp maxima and minima uniformly
separated by 7-8 millisecond intervals was found. 
Here we explore the causes of this newly observed 
phenomenon, and conclude that it is 
either due to a self-organised plasma 
phenomenon or an 
interaction between the plasma and a real-time control system. 
If the maxima are a result of 
``resonant'' frequencies at 
which ELMs can be triggered 
more easily, then future ELM control
techniques can, and probably will, use them.
Either way,  these results demand a deeper understanding of 
the ELMing process.

\end{abstract}

\maketitle

The economically competitive production of 
energy in magnetically confined tokamak plasmas requires
plasmas with high pressures and high energy confinement times. 
In present experiments the majority of high performance plasmas have
edge localised modes (ELMs) \cite{Zohm}, that 
intermittently eject a small fraction of the confined plasma energy
and particles. 
While ELMs are relatively harmless in present machines, 
in larger devices such as ITER \cite{Aymar} they will need to 
be controlled or entirely avoided. 
The presently accepted model for ELMs involves a build-up of pressure
and current at the plasma's edge, that is released by an ELM
\cite{Kamiya},  usually 
presumed to be triggered by a Magnetohydrodynamic instability
\cite{Webster}.  
Here we report results that cannot be explained by this simple picture
alone, 
requiring a revised picture for the causes of ELMs, and suggesting 
new possibilities for how ELMs might be better controlled.

\begin{figure}[htbp!]
\begin{center}
\includegraphics[width=8.5cm]{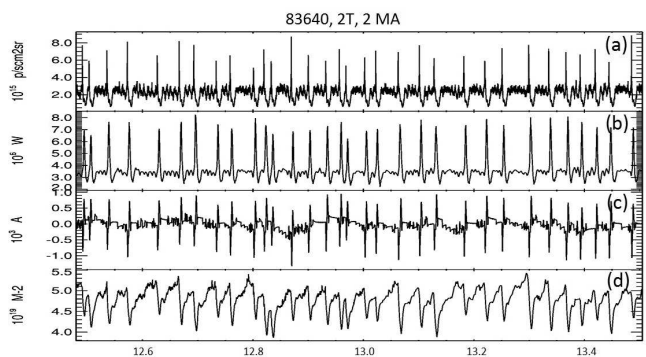}
\end{center}
\vspace{-0.5cm}
\caption{ \label{83640}
Signals from a typical pulse (83640) in the set 83630-83794. 
The signals shown are: Be II radiation (a), the total radiated power (b), 
the current in the EFRA vertical control system (c), and the line
integrated edge density (d).  
The ELMs are associated with a strongly peaked Be II and radiated
power (signals (a) and (b)), a 
rapid response of the vertical control system to keep the plasma
stable (c), and a drop in the edge density (d). 
}
\end{figure}

\begin{figure}[htbp!]
\begin{center}
\Hide{
\includegraphics[scale=0.31]{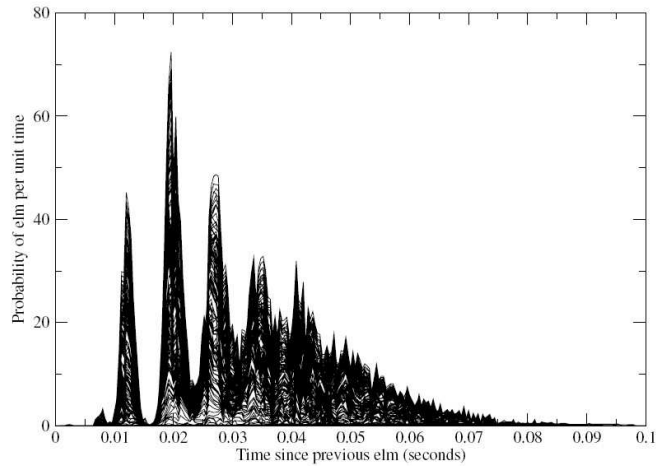}}
\caption{\label{C30}
The ELM waiting time pdf inferred from analysing 120 pulses and
combining the data to form a single pdf. Each line corresponds to data
from an additional pulse.  
}
\end{center}
\end{figure}

During the 2012 experimental campaign at the Joint European Torus
(JET) \cite{Wesson}, the consecutive sequence of pulses 83630-83794 
repeated the same low triangularity 
2T 2MA plasmas with 
approximately 11.5MW neutral beam (NBI) heating and 6 seconds of steady 
high confinement H-mode for 151 good pulses.   
The purpose was to investigate material
migration, fuel retention, and evolution of wall conditioning during
H-mode with the ITER-Like Wall (ILW). 
Excluding pulses that have reports of impurity influxes (``UFOs''),
Nitrogen 
seeding, or any problem preventing them from being steady-state,
leaves 120 nearly-identical pulses  
each 
with approximately 6 seconds of steady type-I H-mode
plasma. 
The Berylium II (527nm) radiation 
was observed at the inner divertor, and the time series of
emissions was analysed, with  
ELMs inferred  from large amplitude signals 
that exceed the average by at least two standard deviations
\cite{Weibull}.   
An example of the signals studied is in figure \ref{83640}.
For each pulse, the number of ELMs with 
waiting times since the previous ELM between time $t$ and $t+0.001$
seconds were counted,  and 
used to form a probability density function (pdf) for
the waiting times between ELMs in the 
9.5-13.5 
second interval.  
Adding together and normalising the 120 pdfs produces figure
\ref{C30}, which combines the data from nearly 15,000 ELMs and 8
minutes of steady state JET plasma time.   

A previous study \cite{Weibull} reported details of 84 high quality
JET datasets for which good agreement was found between the measured
ELM waiting times and a simple but rigorous theoretical model. 
The study was intended to test the theoretical model, which 
applied to ELM waiting time pdfs with a single maximum. 
Consequently the study explicitly excluded datasets with 
more than one maximum. 
In contrast to the pdfs studied in Ref. \cite{Weibull}, figure
\ref{C30} shows  
a sequence of sharp maxima (and minima) separated by 7-8ms
time intervals, corresponding to frequencies of approximately
83, 50, 37, 28, and 24Hz. 
The pdf's variation between maxima and minima is substantial. 
Whereas the
first peak contains 5-10\% of the ELMs, the following minimum  
indicates that there is approximately zero probability of 
observing an ELM at 0.016s after {\it{any}} ELM. 
The structure in the pdf becomes clearer as more data is added, but is
clearly visible once data from 5-6 pulses are combined, corresponding
to about 500 ELMs. 
The same results are found with independent ELM analysis algorithms,
and the phenomenon is not always present in pulses with different 
heating and fueling. 
Therefore we do not think that a diagnostic or analysis
algorithm is incorrectly producing this result, and are confident
that the phenomenon is real. 
Immediate questions are: what is the cause of this
phenomenon? and importantly, do the maxima correspond to physical
resonances at which ELMs could more 
easily be triggered? 

The rest of this article refers to these observed maxima and minima in
the ELM waiting time pdf as ``resonances'', although we 
do not necessarily claim that they are, and
explores the  possible causes of the phenomenon.  
The evidence we will present suggests that the cause is either a
self-organised plasma phenomena, or a control system that is
interacting with the 
plasma in a plasma-dependent way. 
We will conclude by proposing a simple experimental test to decide
whether there are resonant frequencies 
at which ELMs are more easily triggered; this question is key to any
attempts to pace or trigger ELMs in a time-dependent way.

\begin{figure}[htbp!]
\begin{center}
\Hide{
\includegraphics[scale=0.34,angle=-90]{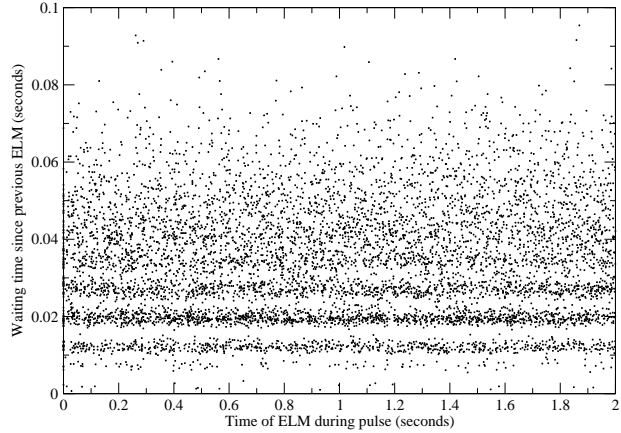}}
\caption{\label{fig2}
The occurence time of ELMs (horizontal axis), is plotted against the
waiting 
time since the previous ELM (vertical axis), for the 83630-83794 pulse
set described in the main text and analysed between 11.5 and 13.5
seconds. The ELM occurence times are  
offset so that the first ELM is at time zero. 
If 
ELMs were being affected by a periodic external system, then we would
expect to see clustering of ELMs into vertical stripes, which we do
not.  
}
\end{center}
\end{figure}

Figure \ref{fig2} shows the occurence times of ELMs (horizontal axis) 
against the waiting time since the previous ELM (vertical axis),  
with the occurence time of ELMs offset so that the first ELM
appears at time $t=0$. 
The waiting time pdf in figure 
\ref{C30} indicates that the waiting times are 
clustered around 0.012, 0.020, 0.028, 0.036, and 0.044 seconds, then more
evenly spread for large time delays. 
This corresponds to the way in which 
ELM waiting times are clustered in 
horizontal stripes in figure \ref{fig2}. 
If the occurence times of successive ELMs are at least approximately
independent (we have found a 
weak negative correlation between successive waiting times), then
beyond the first 0.05 seconds or so, we would not expect to see a
clustering of ELMs with respect to the horizontal time axis. 
The lack of vertical stripes in figure \ref{fig2} is consistent with
this. 
This is a key observation. 
Remember that we have offset the ELM-times so that the first ELM is
at $t=0$, so any pacing with the same frequencies ought to be in
phase. 
If the ELMs had been caused by some 
external influence that was pacing them at the observed but fixed
frequencies, 
then we would expect to retain coherence with respect to the
horizontal ELM time co-ordinate. 
These remarks have been confirmed by Monte Carlo modeling of ELM
occurence times. 
Because the resonances are only observed relative to
consecutive ELMs, we conclude that they are caused  either by a
self-organised 
plasma phenomena, or by an interaction with a real-time plasma control
system. 
It is well known that the real-time vertical control system can
trigger ELMs \cite{Liang}, so it is an obvious potential cause of the
resonances.  
This possibility is considered next.


\begin{figure}[htbp!]
\begin{center}
\Hide{
\includegraphics[scale=0.34,angle=0]{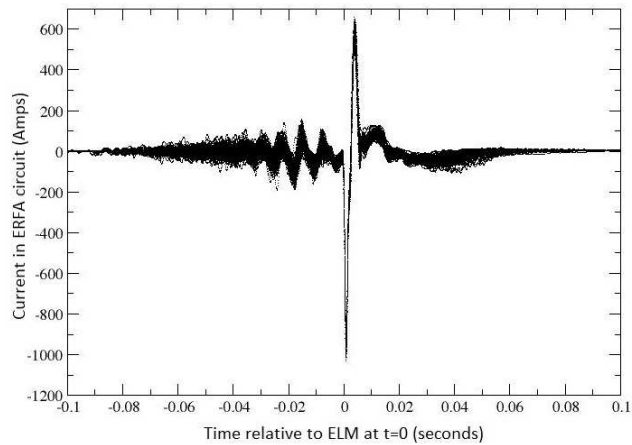}}
\caption{\label{fig3}
The current (Amps) in the ERFA vertical control system measured
between the start of the $(n-1)$th ELM and the start of the $(n+1)$th
ELM, with time offset so  
that the $n$th ELM appears at $t=0$, is averaged over
all ELMs in a pulse and simultaneously plotted for all pulses in our
120 pulse dataset.    
}
\end{center}
\end{figure}

Consider the current flowing in the vertical control system's coils
(the ERFA system \cite{Sartori}), measured from the time of the
$(n-1)$th ELM to the time of the $(n+1)$th ELM, with the time offset
so that time $t=0$ corresponds to the time of the $n$th ELM. 
If we combine and average these over a single pulse, then superimpose
the resulting plots from the 120 pulses in our data set, the result is
figure \ref{fig3}.  
There are a number of striking features. 
First there is a distinctive large-amplitude response of the system
immediately following an ELM, roughly between $t=0$ and $t=0.008$s,
that is the same in different plasmas. 
This response is however known to be 
dependent on the vertical control system settings. 
When JET's carbon plasma facing materials were replaced with the new
ITER-like wall, the vertical control system was modified and
optimised for use with the new wall \cite{Albanese}. 
We have noticed that 
the large amplitude response that immediately follows an ELM is very
different for 
the carbon-wall plasmas, with a large-amplitude
signal that  damps towards zero much less rapidly than in the present
system. 
We have not yet found evidence of resonances in Carbon-wall data. 
Returning to figure \ref{fig3}, as $t$ increases positively the
signals average to zero. 
This indicates that the response of the system to different ELMs is  
out of phase, and differs between ELMs. 
For negative $t$ there is the appearance of an oscillation
in the signal. 
This is a necessary consequence of 
the pdf shown in figure \ref{C30}, that
ensures that the large amplitude signal that
immediately follows the $(n-1)$th ELM is observed predominately at
intervals of 0.012, 0.020, 0.028, 0.036, and 0.044 seconds prior to the
$n$th ELM.  
Because figure \ref{fig3} plots from the start of the $(n-1)$th ELM to
the {\it start} of the $(n+1)$th ELM, the large amplitude signal that
{\it follows}
the start of the $(n+1)$th ELM is not plotted, and consequently similar
oscillations are  not  produced for positive $t$. 
Oscillations are not observed for   
ELMs with waiting times in excess of 0.044 seconds, consistent with
the pdf in figure \ref{C30}.    
These remarks do not rule out a coupling to the vertical control
system, but we have not found clear evidence of one yet. 
The possibility of a coupling between the vertical control
system and the ELMing plasma is  being explored with  
more sophisticated techniques.

A further search of plasmas with the ITER like wall has found that the
resonances are sensitive to heating. 
The plasma parameters of the JET 
H-mode pulses 83393, 83429, and 83593, are equivalent to pulses  
83630-83794, but have only 5-6 MW of NBI heating. 
For these pulses no evidence of resonances has been found. 
For pulse 83155 the heating was increased from the approximately 11.5MW
of neutral beam (NBI) heating in pulses 83630-83794 to 17MW, while the
fueling rate was reduced from approximately 1.15$\times$ 10$^{22}$  to 
0.9 $\times$ 10$^{22}$ particles per second.  
Here again there is no evidence for resonances similar to those
in figure \ref{C30}. 
The sensitivity of the waiting time resonances to the plasma heating
(and possibly also to fueling),   
indicates that they  are either caused by a plasma 
phenomenon, or by an interaction between the plasma and a control
system in real time, and in a way that is sensitive to the plasma's
rate of heating. 
Because a clear observation of resonances requires more ELMs than are
usually present in the steady phase of typical JET H-mode 
plasmas, and even more for higher frequency resonances, it is
presently uncertain how common the ``resonance'' phenomenon is.

The time interval of 0.008s between the observed resonances in figure
\ref{C30} 
could be explained if the plasma was rotating with a frequency of
order 125Hz and interacting with some toroidal asymmetry. 
The rotation rate as measured by the charge exchange diagnositic in
pulses 83630-83794, is greatest in the 
plasma's core, reduces to 
approximately 1kHz at the top of the pedestal, then reduces further
towards the separatrix. 
Unfortunately the uncertainty in the flow measurement increases with
proximity to the separatrix, where the flow rate is likely to be lowest. 
Therefore all we can say with certainty at present is that we do not
know whether the plasma flow in the region between the top of
the pedestal and the 
separatrix could be responsible for the resonances, or not.

From a practical perspective, an important question is: are
there resonant frequencies at which ELMs can be triggered more
easily? 
Fortunately this  can be answered relatively easily without
understanding the cause of the phenomenon, by exploring whether ELMs
in equivalent plasmas can be triggered more (or less) easily with
vertical kicks \cite{Liang} at frequencies of the maxima (or minima)
of the pdf in 
figure \ref{C30}.  
A sensitivity of kick-triggering success to kick frequency was found
in TCV \cite{Degeling}, with similar ranges of kick frequencies
remaining successful (or not), in different plasmas.  
It was suggested that 
the preferred frequencies might be an intrinsic property of the plasma
when it is regarded as a driven dynamical system (\cite{Degeling}, page
1645). 
A similar cause was suggested for the formation of a bimodal ELM
waiting time pdf as gas fueling is systematically increased
\cite{Dendy}. 
Whether this is the correct physical interpretation remains to be
seen, but a carefully designed experiment in conjunction with the
results presented here should conclusively determine whether the
likelihood of triggering an ELM is correlated with the resonances
in figure \ref{C30}. 
Such experiments can provide insights and improve our basic 
understanding of ELMs, possibly leading to an entirely new explanation
for the results presented here, but no-doubt leading 
both directly and indirectly  
to improved methods for plasma control.

The primary experimental results presented in this paper are
unanticipated by theory and, to our knowledge, are not foreshadowed by
previous ELM experiments. A comprehensive understanding of ELM
dynamics is still missing, and it is hoped that the present results
will contribute to its construction. Theory suggests that 
linear instabilities may initiate ELMs after the plasma
current or pressure has passed some threshold value; for a recent
review see \cite{Webster}. 
We note that thresholded instability can give rise to many different
kinds of event time series, spanning the dripping faucet \cite{Faucet}
and sandpile avalanching \cite{Sandpile,Avalanching}. The theoretical
considerations that are candidates for inclusion in such a model span
most of tokamak edge pedestal modelling, and include local turbulence,
transport, and stability, together with the magnetohydrodynamic
character of ELMs and the plasma boundary. We refer to
Refs. \cite{Webster,RMP,EdgeTurb}, and citations therein, for further
discussion of the issues involved and examples.

To conclude, we have found clear examples of plasmas in which the
waiting times between ELMs have preferred frequencies at which ELMs are
more commonly observed.  
This was totally unexpected, and 
is not predicted by present Magnetohydrodynamic  
models for ELMs. 
The phenomenon has been found to depend on the rate of heating, and
the ``resonances'' are observed relative to other ELMs, but not in
absolute time. 
These observations suggest that they are either caused by a 
self organised plasma 
phenomenon or a real-time interaction between the plasma and a control
system. 
We have no clear evidence that they are related to the plasma's
rotation, or to 
an interaction with the vertical control system, but it is 
presently not possible to conclusively rule out 
these possibilities.  
From a practical perspective, 
an important question is whether there are 
frequencies at which ELMs can be more (or less) easily triggered. 
Fortunately this latter question can be answered by using
``vertical kicks'' to explore if ELMs are 
triggered more (or less) easily at resonant (non-resonant) frequencies. 
Because of the relative simplicity but importance of this experiment
for our basic 
understanding of ELMs and ELM control, this is an experiment we
recommend. 
New developments are required to successfully understand and model
this newly observed phenomenon. 
This is likely to include successful modeling of 
the processes by which the post-ELM 
plasma edge reforms prior to successive ELMs, and the 
inclusion of any relevant interactions between the plasma and real-time
control systems. 
Either way, the results here seem to require new lines of research,
and a fresh picture of ELMs and the ELMing process.

\begin{acknowledgments}
{\bf Acknowledgments:} 
This work, part-funded by the European Communities under the
contract of Association between EURATOM/CCFE was carried out within
the framework of the European Fusion Development Agreement.  
For further information on the contents of this paper please contact
publications-officer@jet.efda.org. 
The views and opinions expressed 
herein do not necessarily reflect those of the European Commission. 
This work was also part-funded by the RCUK Energy Programme [grant
number EP/I501045].
\end{acknowledgments}


\begin{thebibliography}{99}

\bibitem{Zohm} H. Zohm, Plasma Physics and Controlled Fusion 38, 105,
  (1996). 

\bibitem{Aymar} R. Aymar et al. for THE ITER TEAM, Plasma Physics and
  Controlled Fusion 44, 519, (2002). 

\bibitem{Kamiya} Kamiya et al., Plasma Physics and Controlled Fusion
  49, S43, (2007). 

\bibitem{Webster} A.J. Webster, Nuclear Fusion 52, 114023, (2012).  

\bibitem{Wesson} J. Wesson {\sl Tokamaks} (Oxford University Press, Oxford,
  1997).  

\bibitem
{Weibull} 
A.J. Webster and R.O. Dendy, Phys. Rev. Lett. 110, 155004, (2013).  

\bibitem{Liang} Y. Liang, Fusion Science and Technology 59, 586,
  (2011). 

\bibitem{Sartori} F. Sartori, G. De Tommasi, F. Piccolo 
IEEE Control Systems Magazine 26, 64-78, (2006).

\bibitem{Albanese} R. Albanese, G. Ambrosino, M. Ariola et al. Fusion
  Engineering and Design 86, 1030, (2011). 

\bibitem{Degeling} A.W. Degeling, Y.R. Martin, J.B. Lister, et
  al. Plasma Phys. Control. Fusion 45, 1637, (2003). 

\bibitem{Dendy} F.A. Calderon, R.O. Dendy, S.C. Chapman,
  A. J. Webster, B. Alper, R. M. Nicol and JET EFDA Contributors,
  Phys. Plasmas, 20, 042306, (2013).


\bibitem{Faucet} B. Ambravaneswaran, S. D. Phillips, and O. A. Basaran
  Phys. Rev. Lett. 85, No. 25, 5332, (2000) 
\bibitem{Sandpile} S C Chapman, R O Dendy and B Hnat,
  Phys. Rev. Lett.  86, 2814 (2001) 
\bibitem{Avalanching} S. Jolliet, Y. Idomura Nucl. Fusion 52, 023026, (2012)
\bibitem{RMP} T E Evans, R A Moyer, K H Burrell, M E Fenstermacher, I
  Joseph et al, Nature Phys. 2, 419 (2006) 
\bibitem{EdgeTurb} W Wan, S E Parker, Y Chen, R J Groebner, Z Yan, A Y
  Pankin, and S E Kruger, Phys. Plasmas 20, 055902 (2013) 

\end{thebibliography}
\end{document}